\documentclass[a4paper]{article}

\usepackage{icrc2013}

\title{The ASTRI SST--2M Prototype: Structure and Mirror}

\shorttitle{ASTRI Prototype: Structure and Mirror}

\authors{Rodolfo Canestrari$^{1}$, Osvaldo Catalano$^{2}$, Mauro Fiorini$^{3}$, Enrico Giro$^{4}$, Nicola La Palombara$^{3}$, Giovanni Pareschi$^{1}$, Luca Stringhetti$^{3}$, Gino Tosti$^{5}$ and Stefano Vercellone$^{2}$ on behalf of the ASTRI collaboration$^{6}$ and Francesco Martelli$^{7}$, Giancarlo Parodi$^{7}$, Pierfrancesco Rossettini$^{8}$ and Raffaele Tomelleri$^{8}$}

\afiliations{
$^1$INAF-Osservatorio Astronomico di Brera\\
$^2$INAF-Istituto di Astrofisica Spaziale e Fisica Cosmica di Palermo\\
$^3$INAF-Istituto di Astrofisica Spaziale e Fisica Cosmica di Milano\\
$^4$INAF-Osservatorio Astronomico di Padova\\
$^5$Universit\`a di Perugia\\
$^6$http://www.brera.inaf.it/astri/\\
$^7$BCV progetti s.r.l.\\
$^8$Tomelleri s.r.l.\\}

\email{rodolfo.canestrari@brera.inaf.it}

\abstract{The next generation of IACT (Imaging Atmospheric Cherenkov Telescope) will explore the uppermost end of the VHE (Very High Energy) domain up to about few hundreds of TeV with unprecedented sensibility, angular resolution and imaging quality.
To this end, INAF (Italian National Institute of Astrophysics) is currently developing a scientific and technological telescope prototype for the implementation of the CTA (Cherenkov Telescope Array) observatory. ASTRI (Astrofisica con Specchi a Tecnologia Replicante Italiana) foresees the full design, development, installation and calibration of a Small Size 4 meter class Telescope. The telescope, named SST--2M, is based on an aplanatic, wide field, double reflection optical layout in a Schwarzschild-Couder configuration.
In this paper we report about the technological solutions adopted for the telescope and for the mirrors. In particular the structural and electro-mechanical design of the telescope and the results on the optical performance derived after the development of a prototype of the segments that will be assembled to form the primary mirror.}


\begin{document}
\maketitle

\section{Introduction}
ASTRI~\cite{astrisys} is a {\it flagship} program financed to INAF by MIUR (Italian Ministry of Education, University and Research) to develop special technologies suitable for the ambitious CTA Observatory~\cite{cta}. CTA will be implemented in two sites, in the northern in the southern hemispheres. It is composed by many tens of telescopes divided in three classes, in order to cover the energy range from a tens of GeV (Large Size Telescope, LST), to a tens of TeV (Medium Size Telescope, MST), and up to 100 TeV and beyond (Small Size Telescope, SST)~\cite{ssticrc}. The SST array will have 70 telescope units and it will be installed only on the southern site.\\
Within this framework, INAF is currently carrying on the development of an end-to-end prototype of SST in a dual-mirror configuration (SST--2M) to be tested under field conditions in Italy at the site of Serra La Nave~\cite{sln}, and scheduled to start data acquisition in late 2014.  All the technological aspects concerning the development of the dual-mirror telescope prototype are covered within ASTRI, including the electro--mechanical mount for the telescope, primary and secondary mirrors, the sensors and the electronics for the camera~\cite{camera}, the Monte Carlo simulations~\cite{montecarlo} and the control and data handling software~\cite{software, data}.  
Telescope mount and mirrors are presented in this paper.
It should be noted that within ASTRI we are also carrying out the design, development and deployment of a mini-array of SST--2M that will be implemented at the CTA southern site and that could constitute the first seed for the main array~\cite{miniarray1, miniarray2, science}.

\section{The telescope mount}
\begin{figure*}[!ht]
  \centering
  \includegraphics[width=0.76\textwidth]{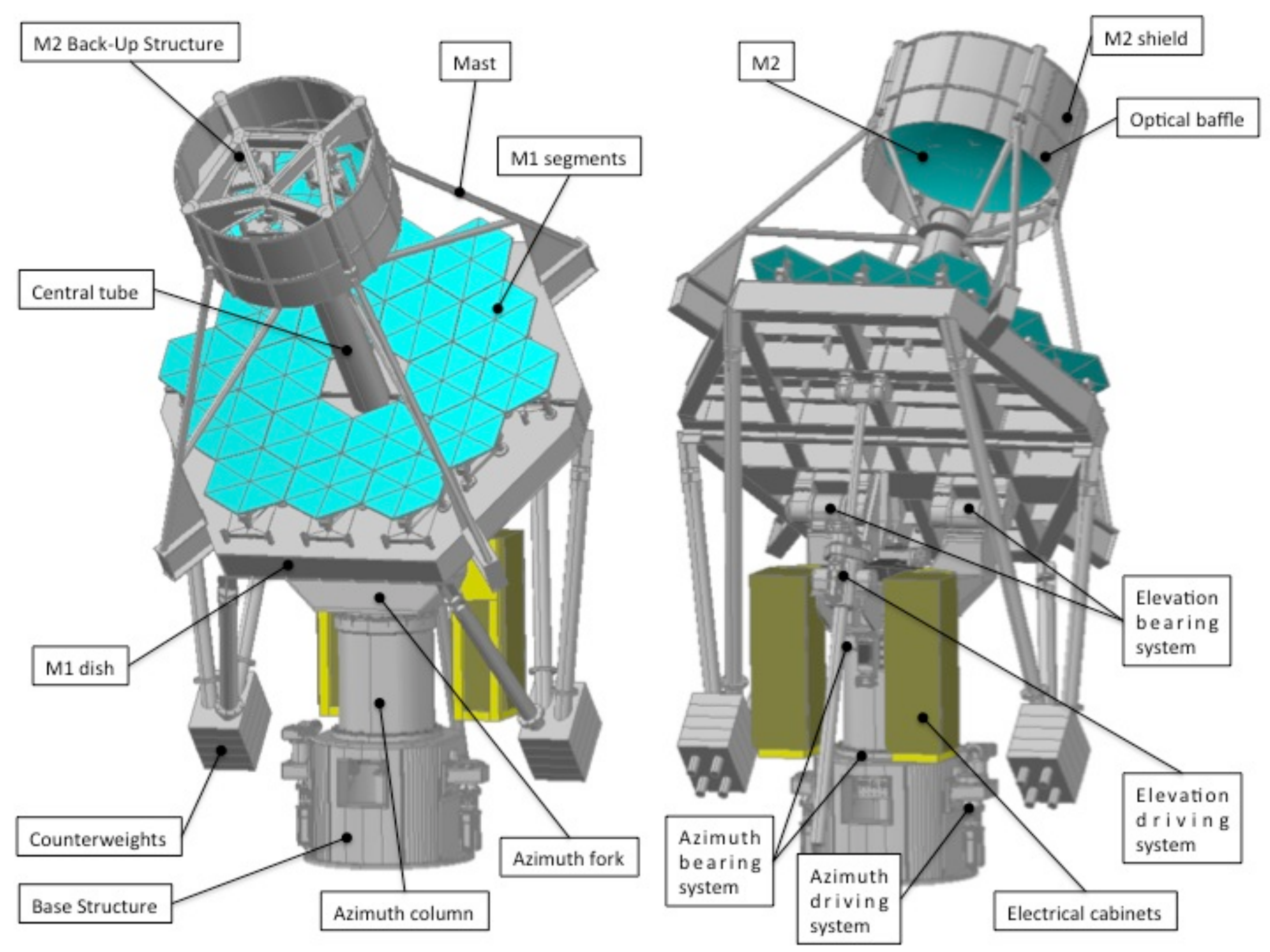}
  \caption{General view and nomenclature of the ASTRI SST--2M telescope structure and electro-mechanical subsystems.}\label{fig:tel-str}
 \end{figure*}

The ASTRI SST--2M telescope is presented in \figurename~\ref{fig:tel-str}. The main parts are the mount, that is composed by the base, the column and the fork, and the optical supporting structure, that is composed by the primary mirror dish (with the mirrors and their supports), the mast with the central tube, the secondary mirror back-up structure (with the mirror and its supports) and the counterweights. \\
The mounting of the telescope is of the alt-azimuthal type. The fork supports the telescope, hosts the elevation subsystems and connects the telescope with the column. The latter hosts the azimuth driving and bearing subsystems. The azimuth axis will admit a useful rotation range between $-270^\circ$ and $+270^\circ$ over a total run of 550$^\circ$; the elevation axis between $-5^\circ$ and $+95^\circ$ over a total run of 110$^\circ$.\\
The M1 dish has a thick ribbed plate to support the 18 mirror segments. It is connected to the mast and, to balance the torque due to its overhang, to two long arms supporting the counterweights. The mast is a slim quadrupode (plus a central tube) with an eccentric symmetry and some radial bracings to improve the bending stiffness. Finally, on top of the mast a structure forms the back-up for the secondary mirror. This also provides the connections to the alignment devices of the mirror and to the covering (shield and optical baffle).\\
The telescope will be made of steel (different grades will be used for the different structural elements) because of a number of advantages (i.e. easy manufacturing, limited thermal gradients, etc.) compared to other structural materials such as Aluminum or Carbon Fiber Reinforced Plastic.\\
The dimensions of the telescope are driven by the optical design already described in the proceeding of the past ICRC~\cite{icrc2011}, and in particular the relative distances between the different optical elements (primary and secondary mirrors and the focal surface). Additional parameters used for the dimensioning take into account the safety of operations around the telescope (e.g. risk of collisions with the counterweights). The dimensioning of the structural components is driven by the project requirements coming from CTA such as the design loads, the tolerances required to maintain the optical and pointing performance. Indeed, the structural checks (e.g. winds, earthquakes, etc.) have been performed in accordance with the most updated international norms and standards (i.e. Eurocodes). The evaluation of the telescope performance in terms of vignetting, optical PSF (Point Spread Function) degradation and pointing precision have been computed, too. However, these kind of performance have been verified by means of ray-tracing the flexures of the telescope structures and mis-alignments/mis-orientations of the optical devices asserted by the structural analyses. \\
Simple dynamical analyses have also been performed in order to investigate the behavior of the telescope. Main results concern the evaluation of the Eigen-frequencies and the expected tracking errors of the telescope. The Eigen-frequencies have a nearly constant behavior with respect to the azimuthal position; the first modes, being the oscillations around the azimuth and elevation axes, can be ascribed to the stiffness of the mechanical components with values in any case higher than 4 Hz. Concerning tracking errors, they return to be of the order of 10 arcsec rms.\\

\subsection{The structure}
\begin{figure}[t]
  \centering
  \includegraphics[width=0.41\textwidth]{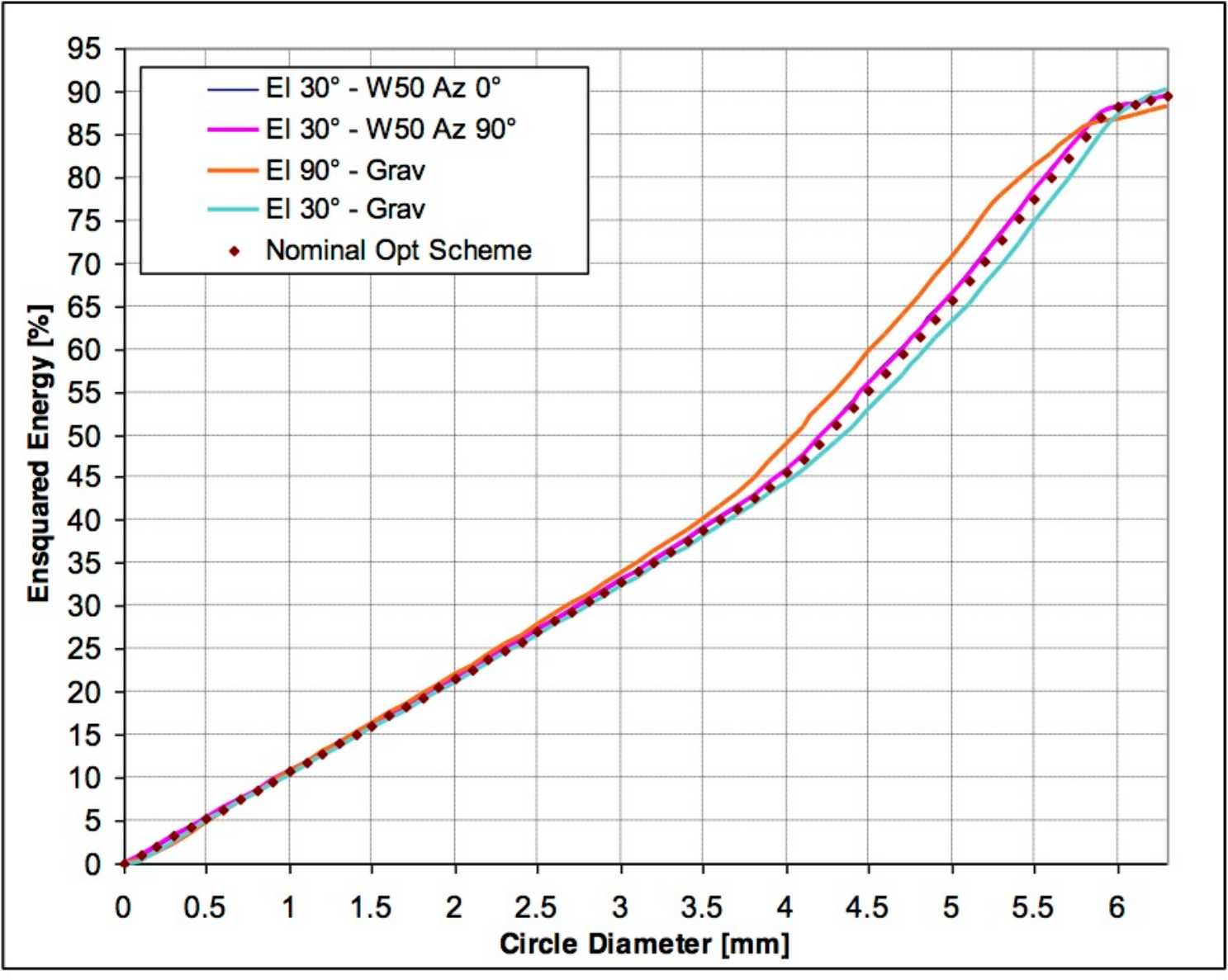}
  \caption{Evaluation of the PSF degradation induced by the telescope structure under different loads.}\label{fig:tel-fem}
 \end{figure}

For each main part of the ASTRI SST--2M structures (i.e. the mast, the M1 and M2 supporting structures) a variety of configurations have been proposed, sketched and evaluated. Through a trade-off activity (typically between performance, mass, easiness of production and assembling) one option is selected for advancing the design. Then a simplified global model of the entire telescope was generated and evaluated. The outcomes of this phase were used to highlight the weak points of the actual design and to allow its refinement to improve the performance and meet the project requirements. Moreover, the forces acting on the mechanical components such as the bearings, the drives, the gears and other devices were evaluated for the proper sizing of the mentioned components. Finally, the very detailed integrated model of the telescope is generated (with about 700.000 degrees of freedom). This model takes into account also the performance specifications of the mechanical components, the mirrors and other devices on board of the telescope (shields, scientific detector, etc.). The full set of loading cases is applied and the comprehensive behavior of the telescope is asserted. All the telescope structural components are modeled and evaluated by means of Finite Elements numerical approach (FE Model and FE Analysis). One of the results achieved, in terms of optical degradation due to mechanical stiffness, is shown in \figurename~\ref{fig:tel-fem}\\
\noindent
{\bf The mast} is composed by an eccentric quadrupode with a radial bracing system connecting the M2 back-up structure, the central tube and the quadrupode legs.. This solution shows increased performance in comparison with simpler configurations. In particular, the quadrupode conferees the adequate rigidity against lateral deformations and, being eccentric, it gives enhanced performance along the elevation axis. Finally, the central tube increases the torsional stiffness and provides the support for the detector.\\
\noindent
{\bf The M1 dish} has a T-ribbed plate structure with a box section in correspondence of the elevation axis in order to increase the torsional stiffness. It is composed of two asymmetric halves connected along the middle plane by means of plugs. The connection line is orthogonal with respect to the elevation plane. Concerning the back side of the dish, the upper half hosts the connection with the linear actuator of the elevation drive system, while the bottom one has the interfaces with the arms of the fork where the elevation axis lies. In the front side, the connections with each mirror segment are hosted. Given the flat geometry of the dish, proper segment spacers provide the gross position of the mirrors.\\
\noindent
{\bf The M2 Back-Up Structure} consists in a proper arrangement of tapered welded I-Beams and a central torsional-stiff triangular prism. The tapered beams connect the four flanged tubes bolted to the mast, while the triangular prism supports the M2 axial load-spreaders. Moreover, two specific box sections provide the connection between the lateral support systems of the secondary mirror and the mast. These box sections also help to increase mast stiffness in the plane of gravity loads.\\

\subsection{The electro--mechanical subsystems}
Three main parts compose the pillar: the base, the column and the fork. The pillar has the main purpose to support and move the telescope main structure (i.e. the M1 dish, the mast and the M2 dish) with the required performance as velocity, acceleration and position accuracy, under the defined loads and environmental conditions. As the classical alt-azimuth mounting has been chosen, the solutions implemented are sketched in \figurename~\ref{fig:tel-str} and described in the following.\\
\noindent
{\bf The Azimuth subsystems} are hosted on the base and column where are respectively located the driving and the bearing mechanisms. The base is a large box fixed on its lower part to the foundation; it hosts the complete driving system for the azimuth axis, composed by two driving chains in a master-slave configuration. Each one is based on a commercial epicycloids gearbox plus a custom designed gearbox. The pinions link the gearboxes to brushless servomotors. In this solution the two motors can work with a preloaded differential torque. For safety reasons both motors are endowed by a braking system that smoothly locks the axis when power fails. In addition, a couple of emergency driving systems are available: a 24 V DC motor with an irreversible gearbox and a shaft for the manual driving. Both the emergency systems are decoupled in operational conditions. The column is composed by two concentric large tubes at the ends of which two ball bearings are located. An adjustable spacer axially preloads them; it has the purpose to avoid the axial backlash and to increase the axial stiffness. The lower bearing and the external tube supports weight of the entire telescope, meanwhile the inner tube (acting as azimuth shaft) and the upper bearing transfer the azimuth movement to the telescope. The positioning is retrieved by means of an absolute angular transducer placed at the lower end of the azimuth shaft, close to the driving system. Finally, an electromechanical stow pin is foreseen at the parking position. It is designed for locking the telescope in position up to the worst load conditions.\\
\noindent
{\bf The Elevation subsystems} are mounted on the fork. The main components are the bearings, the driving and the locking systems. The motion is accomplished by means of a linear actuator. It is composed by a fixed preloaded ball screw (the shaft), a rotating preloaded nut, the gearboxes and the motor. The ball screw is moved up and down by the rotating nut, which is driven by the motorized gearbox set. There is also a couple of joints: one is screwed to the M1 dish and the other one to the fork. The latter also supports the body of the actuator, the gearbox and the motor. Both joints are realized by means of preloaded tapered roller bearings and have their axes parallel to the elevation axis. Concerning the safety aspect, the same solutions adopted for the azimuth axis are duplicated here. The elevation axis lies on the arms of the fork. The shaft is divided into two parts at the end of the arms where two identical bearing system units are located. Each one connects and supports the telescope thought the M1 dish. Each bearing unit is composed by eight tapered roller bearings; they are axially preloaded by means of suitable threaded bushes in order to minimize the backlash. A rotating support is screwed to the backside of the M1 dish. One of the two bearing units hosts the absolute angular encoder to retrieve the positioning. An electromechanical stow pin is foreseen: it allows to lock the telescope in horizontal (the parking position) and zenith positions. Bumpers are also envisaged.\\

\section{The optical surfaces}
\begin{figure}[h]
  \centering
  \includegraphics[width=0.44\textwidth]{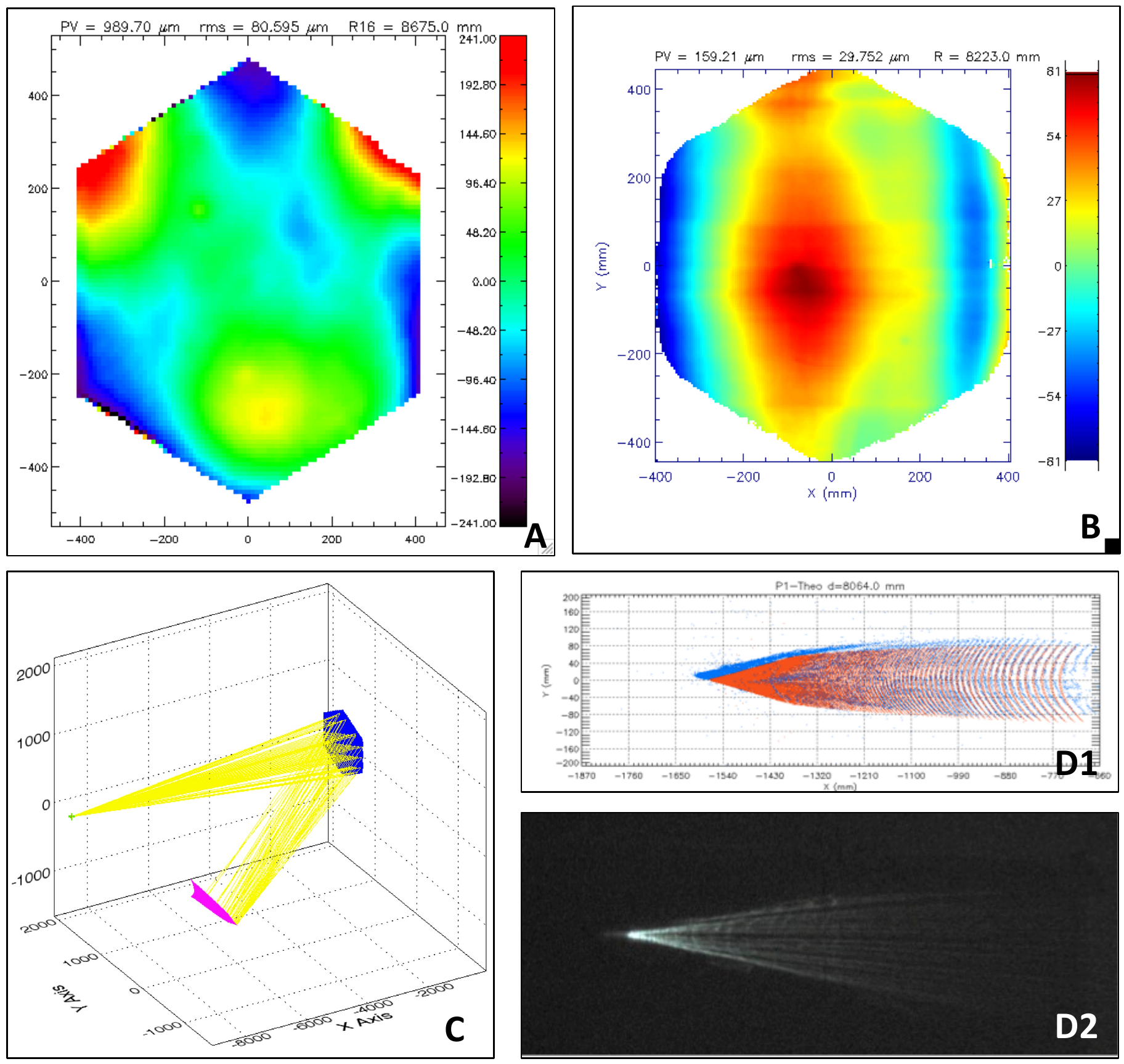}
  \caption{Surface errors of the glass shell (A) and prototype mirror (B). Ray tracing setup (C) and results (D1, D2).}\label{fig:m1}
 \end{figure}

\begin{figure}[h]
  \centering
  \includegraphics[width=0.48\textwidth]{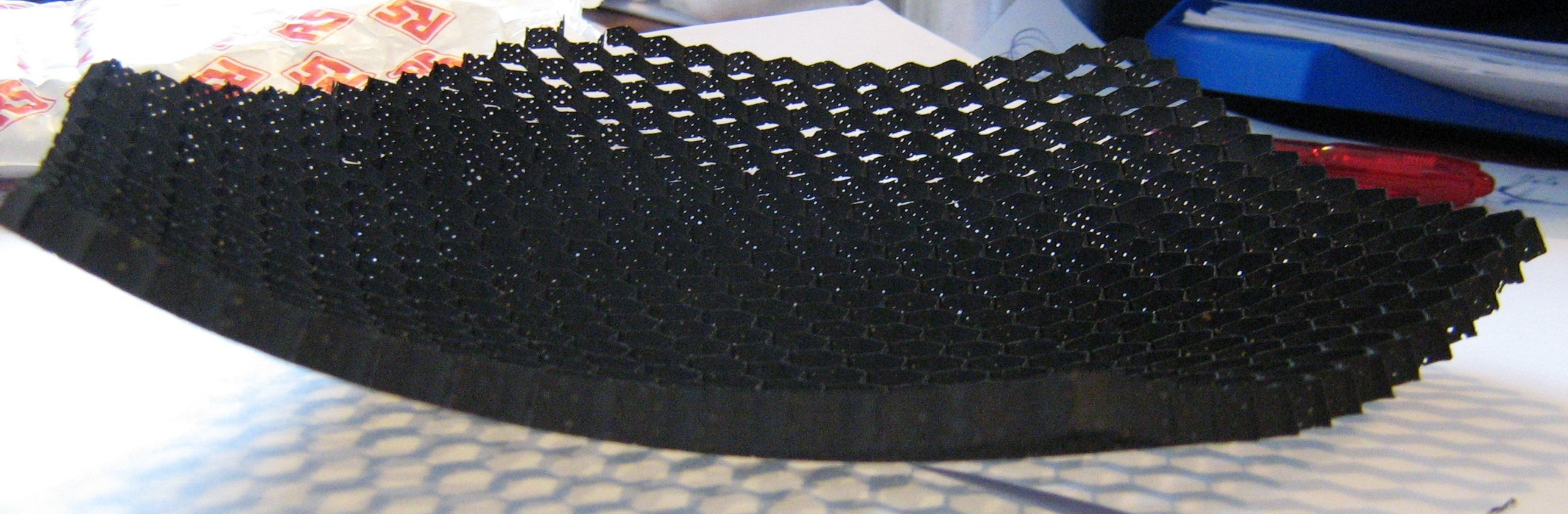}
  \caption{Scaled-down prototype of the honeycomb for M2.}\label{fig:m2}
 \end{figure}

The ASTRI SST--2M telescope implements an aplanatic, wide field, double reflection optical layout known as Schwarzschild-Couder configuration. This concept was initially proposed by Vassiliev et al.~\cite{vlad}. The M1 is tasseled in 18 panels distributed on three concentrical coronae to compose the full $4.3 \;m$ aperture. Since the optical profile follows an aspherical function and each corona has a different radial distance from the telescope vertex, the design of the panels is distinguished for the three cases and consequently different tools are required to produce the whole primary mirror assembly. On the contrary, the M2 is a monolithic element of $1.8 \;m$ in diameter.\\
\noindent
{\bf The technology.} The mirrors (i.e. M1 and M2) are manufactured adopting in both cases a similar technology. The process takes advantage by the replication of the profile from a master shape in order to reproduce several identical mirrors. In particular, the technology used is a modified version of the glass cold slumping process already developed a few years for making the mirrors of the MAGIC--II telescope and now being adopted also for the CTA MST telescopes. The modified version extends the capability of the technology to the manufacturing of mirrors with pronounced curvature and asphericity such those required for the CTA SST telescopes. In a few words, thin glass foils are roughly bent by means of a thermoforming process. The shells are then assembled in a stiff and lightweight sandwich structure by means of an honeycomb core. This last step of the process is done by precisely forming the shells over a master shape. It is performed at room temperature using suitable materials and tools. 
 \begin{figure}[!t]
  \centering
  \includegraphics[width=0.5\textwidth]{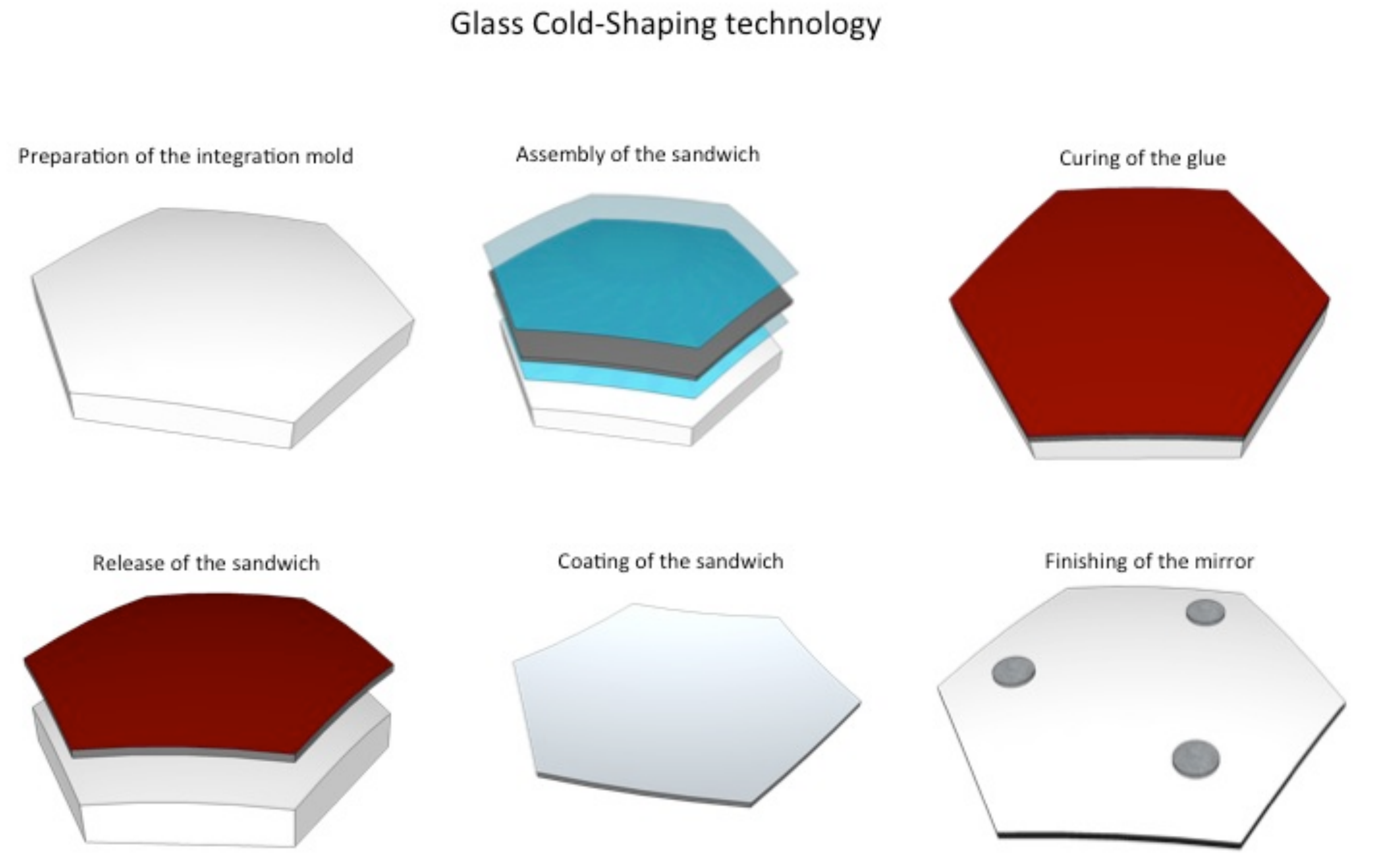}
  \caption{Manufacturing process. Top-line: integration mold, sandwich preparation, polymerization of the glue. Bottom-line: mirror release, coating, finishing.}\label{fig:cold}
 \end{figure}
A simple sketch of the process can be found in \figurename~\ref{fig:cold}, while more details and early results are presented in a number of publications~\cite{cold1, cold2, cold3, hot1}. At the end, high reflective coating is applied as described by Bonnoli et al.~\cite{coating}.\\
\noindent
{\bf Primary mirror.} The opto--mechanical drawings of the replication molds are in agreement with the  optical requirements of the mirror segments. After the machining, the molds shape quality are evaluated by means of coordinate measuring machine and analyzed to find the shape errors. They are characterized, with respect to the nominal profile, by an error  $<30 \; \mu$m rms. Similar measures and evaluations have been conducted on the thermal formed glass shells in order to evaluate the presence of local shape errors at medium/high frequencies that can be only partially corrected by the cold slumping process. In this respect, the thermally pre-shaped shells are characterized by an error $<90 \; \mu$m rms. Finally, some full prototypes have been realized and measured. The segments show an almost perfect match with the mold profile since the shape errors are $<30 \; \mu$m rms. Afterwards, by ray-tracing analysis we have simulated the optical path of the incoming photons for the prototypes realized so far. The surface quality achieved returns to be compliant with the optical requirements of the project. In \figurename~\ref{fig:m1} we summarize the results.\\
\noindent
{\bf Secondary mirror.} Some dedicated activities are ongoing in parallel to those reported for the primary mirror but with further adaptations to account for the peculiarity of the dimensions and curvature. As an example, \figurename~\ref{fig:m2} shows an early prototype of the carbon fibre honeycomb with built-in curvature that is under development.

\section{Conclusions}
In this paper we have introduced the ASTRI project and its main activities toward the development of a prototype for an innovative Small Size Telescope for the CTA Observatory. Details have been presented on some relevant aspects related to the telescope structure and mirrors. In particular, the structural design and the electro-mechanical subsystems that will be implemented into the prototype. Moreover, some highlights on the mirrors technology have been presented, too. The manufacturing and installation of the telescope will take place over the forthcoming months at the Serra La Nave site on the Etna mount in Italy.

\vspace*{0.5cm}
\noindent
\footnotesize{{\bf Acknowledgment: }{This work was supported by the ASTRI ÒFlagship ProjectÓ financed by the Italian Ministry of Education, University, and Research (MIUR) and led by the Italian National Institute of Astrophysics (INAF). We also acknowledge partial support by the MIUR Bando PRIN 2009.}}

\end{document}